\documentclass[%
 aip,
 cha,
 amsmath,amssymb,
 reprint,%
 floatfix,%
]{revtex4-1}

\usepackage{graphicx}
\usepackage{float}
\usepackage{dcolumn}
\usepackage{bm}
\usepackage{booktabs}
\usepackage[utf8]{inputenc}
\usepackage[T1]{fontenc}
\usepackage{mathptmx}
\usepackage{etoolbox}
\usepackage[section]{placeins}

\makeatletter
\def\@email#1#2{%
 \endgroup
 \patchcmd{\titleblock@produce}
  {\frontmatter@RRAPformat}
  {\frontmatter@RRAPformat{\produce@RRAP{*#1\href{mailto:#2}{#2}}}\frontmatter@RRAPformat}
  {}{}
}%
\makeatother

\begin{document}

\preprint{AIP/123-QED}

\title[Biharmonic coupling in frequency-weighted Kuramoto oscillators]
{Effect of biharmonic coupling in frequency-weighted Kuramoto oscillators}

\author{R.~Senthamizhan}
\affiliation{Department of Physics, Centre for Nonlinear Science and Engineering,
School of Electrical and Electronics Engineering,
SASTRA Deemed University, Thanjavur 613\,401, Tamil Nadu, India}

\author{R.~Gopal}
\affiliation{Department of Physics, Centre for Nonlinear Science and Engineering,
School of Electrical and Electronics Engineering,
SASTRA Deemed University, Thanjavur 613\,401, Tamil Nadu, India}

\author{V.~K.~Chandrasekar}
\email{gopalphysics@gmail.com, chandru25nld@gmail.com}
\affiliation{Department of Physics, Centre for Nonlinear Science and Engineering,
School of Electrical and Electronics Engineering,
SASTRA Deemed University, Thanjavur 613\,401, Tamil Nadu, India}

\date{\today}

\begin{abstract}
We investigate the collective dynamics of globally coupled phase
oscillators with frequency-weighted biharmonic coupling. Although
frequency-weighted and biharmonic extensions of the Kuramoto model have
each been studied independently, their combined effect has remained
unexplored. By combining numerical branch continuation with analytical
stability analysis, we construct the complete phase diagram in the
two-parameter coupling plane. The competition between the first- and
second-harmonic interactions gives rise to three distinct collective
states: the incoherent state (IC), the antipodal multibranch state
(APMS), and the symmetry-broken multibranch state (SBMS). These states
are separated by three bistable regions and a narrow tristable region in
which all three states coexist as stable attractors. The transitions
from the incoherent state to both ordered states are discontinuous and
hysteretic, reflecting the explosive synchronization induced by
frequency weighting, while the transition between the APMS and SBMS is
also abrupt. Analytically, we derive the linear instability thresholds
of the incoherent state together with the existence and local stability
conditions for the SBMS, and show that the theoretical predictions are
in good agreement with extensive numerical simulations.
\end{abstract}

\maketitle

\begin{quotation}
Synchronization of large populations of coupled oscillators is a
recurring theme across physics, biology, and engineering, with the
Kuramoto model serving as its canonical mathematical framework. Two
physically motivated extensions of the classical model have each been
shown to generate striking departures from the standard synchronization
scenario. The first is frequency-weighted coupling, in which faster
oscillators exert proportionally stronger influence on the mean field,
leading to explosive synchronization. The second is biharmonic coupling,
where the interaction function contains a $\pi$-periodic second-harmonic
component, giving rise to multibranch phase locking. Although these two
mechanisms have been extensively studied independently, their combined
effect has remained unexplored. Here we investigate the interplay
between frequency weighting and biharmonic coupling and show that it
organizes the collective dynamics into an incoherent state (IC), an
antipodal multibranch state (APMS), and a symmetry-broken multibranch
state (SBMS), separated by multiple bistable and a tristable regions. The transitions from incoherence are explosive,
whereas the competition between the two synchronized states produces a
broad coexistence regime. Our analysis combines large-scale numerical
branch continuation with analytical conditions for the instability of
the incoherent state and the existence and stability of the SBMS. These
results enrich the phenomenology of mean-field oscillator ensembles and
may be relevant to systems, such as electrochemical oscillator arrays
and biological neural circuits, in which higher-harmonic interactions
and frequency-dependent coupling strengths coexist.
\end{quotation}

\section{Introduction}\label{sec:intro}

Synchronization of coupled oscillators underpins the functioning of a
wide range of natural and engineered systems, ranging from power grids
and micro-electromechanical systems (MEMS) to neuronal populations in
the brain. In power grids, generators must maintain phase coherence to
ensure stable power transmission,\cite{motter2013spontaneous} while
synchronized MEMS oscillator arrays exploit collective motion to enhance
signal strength and reduce phase
noise.\cite{zhang2015synchronization} Conversely, excessive
synchronization can be detrimental. Pathological beta-band synchrony in
the basal ganglia is associated with the motor symptoms of Parkinson's
disease,\cite{hammond2007pathological} and hyper-synchronous neural
activity is a defining characteristic of epileptic
seizures.\cite{jiruska2014modern} Understanding the mechanisms that
promote, suppress, or reshape synchronization therefore remains a
central problem in nonlinear science.

The Kuramoto model, introduced in
Ref.~\onlinecite{kuramoto1975self}, has become the canonical framework
for studying synchronization in populations of coupled phase
oscillators. Its mathematical tractability, which admits a mean-field
reduction and, for certain frequency distributions, exact
self-consistency equations,\cite{ott2008,ott2009} has enabled detailed
analytical studies of synchronization onset and the stability of
coherent
states.\cite{strogatz2000kuramoto,acebron2005kuramoto,rodrigues2016kuramoto}
To capture the richer dynamical behavior observed in realistic oscillator
networks, numerous extensions of the classical Kuramoto model have been
proposed over the past decades.

One extension of particular relevance to the present work is
\emph{frequency-weighted coupling}, in which the effective coupling
strength experienced by each oscillator is modulated by a function of
its natural frequency. In this framework, oscillators with larger
natural frequencies experience stronger effective coupling to the
mean field. This seemingly simple modification leads to qualitative
departures from the classical Kuramoto phenomenology, including
hysteretic (explosive) synchronization
transitions,\cite{xu2016synchronization,gmezgardees2011} suppression of
synchronization by repulsive
coupling,\cite{xu2016dynamics} the emergence of quantized
time-dependent cluster states known as Bellerophon
states,\cite{bi2016coexistence} and two-step synchronization
transitions.\cite{ameli2024twostep} The explosive character of these
transitions is particularly noteworthy because it resembles abrupt
synchronization phenomena observed experimentally in mercury
beating-heart oscillators,\cite{kumar2015explosive} turbulent reactive
flow systems,\cite{joseph2024explosive} and coupled photochemical
oscillators.\cite{taylor2008clusters}

A second extension that motivates the present study is
\emph{biharmonic coupling}, in which the sinusoidal interaction of the
classical Kuramoto model is replaced by the first two Fourier modes of
the phase-difference coupling
function.\cite{hansel1993} The second-harmonic interaction introduces
an intrinsic $\pi$-periodicity that fundamentally enriches the
synchronization landscape. It allows oscillators with similar natural
frequencies to lock at multiple distinct phase differences, generating
multibranch synchronized states beyond a critical coupling
strength.\cite{skardal2011,komarov2013multiplicity} Furthermore, the
biharmonic Kuramoto model provides a minimal setting for
self-consistent partial synchrony, namely collective states that are
neither fully coherent nor completely
incoherent.\cite{clusella2016minimal} The physical relevance of
biharmonic interactions is supported by experiments on electrochemical
oscillator arrays, where higher-order coupling naturally arises from
the interplay of linear and quadratic
feedback.\cite{rusin2009framework,kori2014} More generally, since
interaction functions in real oscillator systems are rarely purely
sinusoidal,\cite{daido1996onset,bick2017} biharmonic coupling provides
a natural extension of the classical Kuramoto framework.

Despite the significant progress made on frequency-weighted and
biharmonic extensions independently, their combined influence on
collective synchronization has remained largely unexplored. This
represents an important gap because both mechanisms naturally arise in
physical oscillator systems and affect synchronization through distinct
processes. Frequency weighting differentiates the effective coupling
strengths of oscillators according to their natural frequencies and
promotes abrupt synchronization transitions, whereas biharmonic
coupling introduces competing phase-locking configurations, multibranch
synchronization, and partial synchrony.\cite{boccaletti2016explosive}
Their interplay is therefore expected to generate synchronization
patterns and transition scenarios that cannot be inferred from either
mechanism alone. The present work addresses this problem by
investigating a globally coupled population of oscillators with
frequency-weighted biharmonic interactions.

Our analysis reveals a rich phase diagram in the first- and
second-harmonic coupling parameter space $(K_1,K_2)$, containing
multiple synchronized states and hysteretic transition boundaries,
obtained through a combination of analytical theory and systematic
numerical continuation. The main contributions of this work are
summarized as follows.

\begin{enumerate}
\item We identify a collective state, the \emph{antipodal multibranch
state} (APMS), consisting of a stationary, frequency-sorted,
four-cluster synchronized state generated by dominant second-harmonic
coupling. The APMS combines the antipodal phase organization of
$\pi$-states with multibranch synchronization and demonstrates that
such a state can emerge in a unimodal oscillator population solely
through the interaction structure.

\item We characterize a \emph{symmetry-broken multibranch state}
(SBMS), a stationary two-cluster synchronized state in which the
reflection symmetry between oscillators with opposite natural
frequencies is spontaneously broken. Unlike the APMS, the two clusters
are separated by a phase difference that deviates from $\pi$, resulting
in finite first- and second-harmonic coherence.

\item We derive explicit analytical thresholds for the linear
instability of the incoherent state, obtaining the critical couplings
$K_1^{(c)}=K_2^{(c)}=4$ for the Lorentzian frequency distribution with
unit half-width.

\item We derive the existence relation and local-stability condition
for the SBMS and combine these analytical predictions with numerical
continuation to construct the phase diagram containing mono-stable,
bistable, and tristable synchronization regimes.
\end{enumerate}

The remainder of the paper is organized as follows.
Section~\ref{sec:model} introduces the frequency-weighted biharmonic
Kuramoto model, the order parameters, and the mean-field formulation.
Section~\ref{sec:results} presents the numerical characterization of the
collective states together with one-parameter continuation studies.
Section~\ref{sec:4} develops the analytical theory, including the linear
stability analysis of the incoherent state and the existence and
stability conditions for the symmetry-broken multibranch state (SBMS).
Section~\ref{subsec:pd_overview} presents the two-parameter coupling
phase diagram, combining the numerical results with the analytical
stability boundaries to identify the mono-stable, bistable, and
tristable synchronization regimes. Finally,
Section~\ref{sec:conclusion} summarizes the main findings.

\section{Model and order parameters}\label{sec:model}

\subsection{Microscopic model}\label{subsec:micromodel}

We consider a population of $N$ globally coupled phase oscillators with
phases $\theta_i\in[0,2\pi)$ and natural frequencies
$\omega_i\in\mathbb{R}$, where $i=1,\ldots,N$. The dynamics of the
frequency-weighted biharmonic Kuramoto model are governed by
\begin{equation}
\label{eq:model}
\dot{\theta}_i=\omega_i+
\frac{|\omega_i|}{N}\sum_{j=1}^{N}
\left[
K_1\sin(\theta_j-\theta_i)
+
K_2\sin2(\theta_j-\theta_i)
\right],
\end{equation}
where $K_1$ and $K_2$ denote the coupling strengths of the first- and
second-harmonic interactions, respectively.

The model combines two mechanisms that govern the collective dynamics.
The first-harmonic interaction favors conventional phase synchronization,
where oscillators tend to lock with a single preferred phase difference.
In contrast, the second-harmonic interaction introduces an additional
interaction with period $\pi$, allowing oscillators to synchronize at
multiple phase differences separated by $\pi$ \cite{gong2019}. The
frequency-weighting factor $|\omega_i|$ further modifies the interaction
strength by assigning stronger coupling to oscillators with larger
natural frequencies. Consequently, oscillators with different natural
frequencies contribute unequally to the mean-field interaction, leading
to rich collective behavior including abrupt synchronization
transitions, multistability, and multibranch synchronized states.

Throughout this work, the natural frequencies are drawn from the
Lorentzian distribution
\begin{equation}
\label{eq:lorentz}
g(\omega)=\frac{1}{\pi}\frac{1}{1+\omega^2},
\qquad
\int_{-\infty}^{\infty}g(\omega)\,d\omega=1,
\end{equation}
which is symmetric about $\omega=0$, i.e.,
\begin{equation}
g(\omega)=g(-\omega).
\label{eq:gsym}
\end{equation}

The model possesses the global rotational symmetry common to
Kuramoto-type systems. Since the coupling depends only on phase
differences, Eq.~(\ref{eq:model}) is invariant under the global phase
transformation
\begin{equation}
\theta_i\rightarrow\theta_i+\alpha,
\qquad
i=1,\ldots,N,
\label{eq:rotation}
\end{equation}
where $\alpha$ is an arbitrary constant. Consequently, only the
relative phase differences determine the dynamics, and the synchronized
state is defined only up to an arbitrary overall phase shift.

More importantly for the present work, the symmetry of the natural
frequency distribution allows synchronized solutions satisfying the
reflection relation
\begin{equation}
\theta(\omega)=-\theta(-\omega),
\label{eq:reflection}
\end{equation}
which relates oscillators with opposite natural frequencies. This
reflection-symmetric organization plays a central role in the collective
states discussed in this work.

The second-harmonic interaction further possesses an intrinsic
$\pi$-periodicity. Defining the phase difference
\begin{equation}
\Delta\theta_{ij}=\theta_j-\theta_i,
\end{equation}
the first- and second-harmonic interactions satisfy
\begin{equation}
\sin(\Delta\theta_{ij}+\pi)
=
-\sin(\Delta\theta_{ij}),
\end{equation}
and
\begin{equation}
\sin\!\left[2(\Delta\theta_{ij}+\pi)\right]
=
\sin(2\Delta\theta_{ij}),
\label{eq:piperiod}
\end{equation}
respectively. Thus, while the first-harmonic interaction changes sign
under a phase shift of $\pi$, the second-harmonic interaction remains
invariant. Consequently, when the second-harmonic interaction dominates,
oscillators can synchronize into two equivalent phase configurations
separated by $\pi$, giving rise to antipodal cluster states.

Together, the reflection symmetry associated with the symmetric
frequency distribution and the $\pi$-periodicity of the second-harmonic
interaction provide the structural basis for the antipodal multibranch
state (APMS). When the first-harmonic interaction becomes dominant, the
reflection symmetry of the synchronized solution is spontaneously
broken, giving rise to the symmetry-broken multibranch state (SBMS).
The role of these symmetry properties in organizing the various
collective states is examined in Sec.~III.

\subsection{Order parameters}\label{subsec:orderparams}

The macroscopic state of the ensemble is characterized by the Daido
order parameters\cite{daido1996onset}
\begin{equation}\label{eq:daido}
  Z_m(t)
    \equiv r_m(t)\, e^{i\psi_m(t)}
    = \frac{1}{N}\sum_{j=1}^{N} e^{i m \theta_j(t)},
  \qquad m = 1,\,2,
\end{equation}
where $r_m \in [0,1]$ is the amplitude and
$\psi_m \in [0,2\pi)$ is the phase of the $m$th-harmonic order
parameter. The first-harmonic amplitude $r_1$ measures conventional
phase coherence, while the second-harmonic amplitude $r_2$ detects
$\pi$-periodic clustering: $r_2 \approx 1$ indicates that the
oscillators form antipodal pairs separated by a phase difference
of $\pi$.

\subsection{Mean-field equations}\label{subsec:meanfield}

Using the order parameters~\eqref{eq:daido}, the all-to-all sum in
Eq.~\eqref{eq:model} can be written in mean-field form,
\begin{equation}\label{eq:model_full}
  \dot{\theta}_i
    = \omega_i
    + |\omega_i|\!\left[
        K_1\, r_1 \sin(\psi_1 - \theta_i)
      + K_2\, r_2 \sin(\psi_2 - 2\theta_i)
      \right],
\end{equation}
in which each oscillator is driven by the collective fields $r_m$ and
$\psi_m$ rather than by the individual phases of its neighbours.

In the thermodynamic limit $N \to \infty$, the microscopic state is
replaced by a phase density $f(\theta,\omega,t)$ normalized so that
$\int_0^{2\pi} f(\theta,\omega,t)\,d\theta = g(\omega)$. The order
parameters then become
\begin{equation}\label{eq:daido_cont}
  Z_m(t)
    = \int_{-\infty}^{\infty}\!\int_0^{2\pi}
        e^{im\theta}\, f(\theta,\omega,t)\,d\theta\,d\omega,
\end{equation}
with $m=1,2$, and the density obeys the continuity equation
\begin{equation}\label{eq:cont_eq}
  \partial_t f + \partial_\theta\bigl(v(\theta,\omega,t)\,f\bigr) = 0,
\end{equation}
where the single-oscillator velocity field is
\begin{equation}\label{eq:velocity}
  v(\theta,\omega,t)
    = \omega + |\omega|\!\left[
        K_1\, r_1 \sin(\psi_1 - \theta)
      + K_2\, r_2 \sin(\psi_2 - 2\theta)
      \right].
\end{equation}

To eliminate the global rotational drift, we fix a co-rotating gauge
in which $\psi_1 = \psi_2 = 0$, so that the order
parameters become purely real, $r_m = \langle \cos m\theta \rangle$.
Introducing $\sigma = \operatorname{sgn}(\omega) \in \{+1,-1\}$,
the velocity field~\eqref{eq:velocity} reduces to
\begin{equation}\label{eq:vfield_gauge}
  \dot{\theta}
    = |\omega|\bigl(\sigma
        - K_1 r_1 \sin\theta
        - K_2 r_2 \sin 2\theta\bigr).
\end{equation}
Locked oscillators satisfy $\dot{\theta}=0$, and their phases are
determined by the fixed-point condition
\begin{equation}\label{eq:lock}
  \sigma = K_1 r_1 \sin\theta + K_2 r_2 \sin 2\theta.
\end{equation}
Equations~\eqref{eq:model_full}--\eqref{eq:lock} provide the starting
point for the analytical results presented in Sec.~IV, where we perform
the linear stability analysis and derive the existence and stability
conditions for the SBMS.
\FloatBarrier

\section{Numerical results}\label{sec:results}

We numerically integrated the frequency-weighted biharmonic model, Eq.~\eqref{eq:model_full}, using an adaptive Runge--Kutta--Fehlberg integration scheme. The natural frequencies were drawn from the Lorentzian distribution given in Eq.~\eqref{eq:lorentz}, and the initial phases were chosen randomly from a uniform distribution over the interval $[-\pi,\pi)$. Unless otherwise stated, all simulations were performed with $N=2000$ oscillators.  The order parameters $r_1$ and $r_2$ were obtained by averaged after discarding the initial transients.  In the following, we first characterize the three collective states supported by the model. We then examine the transitions between these states by varying the coupling strengths adiabatically in both the forward and reverse directions.

\subsection{Collective states}\label{subsec:states}

The model described by Eq.~\eqref{eq:model_full} supports three distinct collective states: the incoherent state (IC), the antipodal multibranch state (APMS), and the symmetry-broken multibranch state (SBMS). These states are clearly distinguished by their phase distributions and the corresponding values of the Daido order parameters, $r_1$ and $r_2$.

\begin{figure}[tb]
\centering
\includegraphics[width=\linewidth]{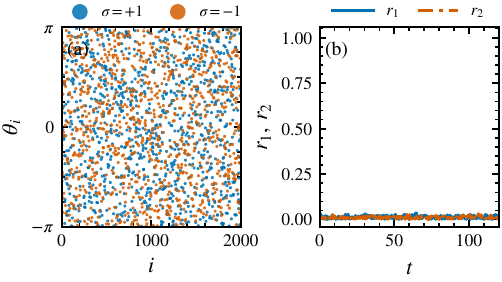}
\caption{Incoherent state (IC). (a) Snapshot of the oscillator phases $\theta_i$ as a function of the oscillator index $i$, with colors indicating the sign ($\sigma_i=\pm1$) of the natural frequencies. (b) Time evolution of the first- and second-harmonic order parameters, $r_1(t)$ and $r_2(t)$, showing that both fluctuate around zero, indicating the absence of collective synchronization. The parameter values are $K_1=-5.0$ and $K_2=-5.0$.}
\label{fig:ic}
\end{figure}

The incoherent state (IC) is characterized by a uniform distribution of oscillator phases over the interval $[-\pi,\pi)$, indicating the absence of collective synchronization. Figure~\ref{fig:ic}(a) shows that the oscillators are randomly distributed around the unit circle without forming any coherent clusters. Consequently, both the first- and second-order parameters fluctuate around zero, as shown in Fig.~\ref{fig:ic}(b), confirming the absence of phase ordering.

\begin{figure}[tb]
\centering
\includegraphics[width=\linewidth]{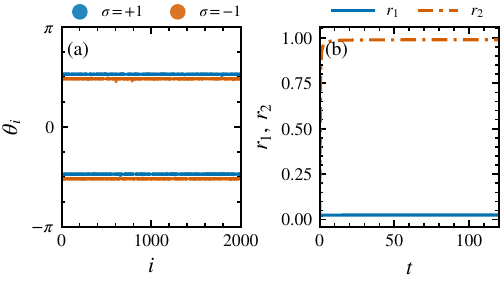}
\caption{Antipodal multibranch state (APMS). (a) Snapshot of the oscillator phases $\theta_i$ as a function of the oscillator index $i$, with colors indicating the sign ($\sigma_i=\pm1$) of the natural frequencies. The oscillators form four synchronized clusters arranged as two antipodal pairs. (b) Time evolution of the first- and second-harmonic order parameters, $r_1(t)$ and $r_2(t)$, showing $r_1\approx0$ and $r_2\approx1$, characteristic of second-harmonic synchronization. The parameter values are $K_1=-5.0$ and $K_2=7.0$.}
\label{fig:apms}
\end{figure}

The antipodal multibranch state (APMS) is characterized by the formation of four synchronized clusters arranged as two antipodal pairs, as shown in Fig.~\ref{fig:apms}(a). Owing to the symmetric Lorentzian frequency distribution, oscillators with positive and negative natural frequencies form two mirror-image branches satisfying the reflection symmetry $\theta(+\omega)=-\theta(-\omega)$. Furthermore, within each frequency branch ($\sigma=\pm1$), the synchronized oscillators split into two clusters separated by a phase difference of $\pi$, giving rise to an antipodal cluster configuration. Because the two antipodal clusters contribute equally but with opposite signs to the first-harmonic order parameter, their contributions cancel exactly, resulting in $r_1=0$. In contrast, both clusters contribute constructively to the second-harmonic order parameter, producing a large value of $r_2$, as shown in Fig.~\ref{fig:apms}(b). Thus, the APMS represents a stationary second-harmonic synchronized state that preserves both the reflection symmetry between positive- and negative-frequency oscillators and the antipodal cluster arrangement within each frequency branch.

For the ideal APMS, where $r_1=0$, the locking condition in Eq.~\eqref{eq:lock} depends only on the second-harmonic coupling. The corresponding self-consistency relation reduces to
\begin{equation}\label{eq:r2_apms}
  r_2^4 - r_2^2 + \frac{1}{K_2^2} = 0,
  \qquad
  r_2(K_2) = \sqrt{\frac{1+\sqrt{1-4/K_2^2}}{2}},
\end{equation}

which admits real solutions only for $|K_2|\ge2$. This analytical result agrees well with the numerical observations and confirms that the APMS is sustained entirely by the second-harmonic interaction.

\begin{figure}[tb]
\centering
\includegraphics[width=\linewidth]{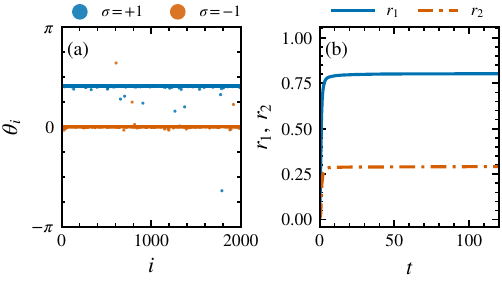}
\caption{Symmetry-broken multibranch state (SBMS). (a) Snapshot of the oscillator phases $\theta_i$ as a function of the oscillator index $i$, with colors indicating the sign ($\sigma_i=\pm1$) of the natural frequencies. The oscillator population forms two synchronized clusters with broken reflection symmetry and without the antipodal cluster arrangement observed in the APMS. (b) Time evolution of the first- and second-harmonic order parameters, $r_1(t)$ and $r_2(t)$, showing finite values of both order parameters. The parameter values are $K_1=5.0$ and $K_2=-5.0$}
\label{fig:sbms}
\end{figure}

As the first-harmonic interaction becomes increasingly dominant relative
to the second harmonic, the system supports a distinct synchronized
state, namely the symmetry-broken multibranch state (SBMS). The
corresponding phase distribution is shown in
Fig.~\ref{fig:sbms}(a). Unlike the APMS, which consists of four
antipodally arranged clusters, the SBMS is organized into two
phase-locked clusters, one predominantly composed of oscillators with
positive natural frequencies ($\sigma=+1$) and the other of oscillators
with negative natural frequencies ($\sigma=-1$). The reflection
symmetry, $\theta(+\omega)=-\theta(-\omega)$, is no longer preserved,
and simultaneously the antipodal ($\pi$-periodic) cluster arrangement is
lost. Consequently, the exact cancellation responsible for $r_1=0$ in
the APMS no longer occurs, and both $r_1$ and $r_2$ attain finite
values, as shown in Fig.~\ref{fig:sbms}(b).

Although a small number of oscillators appear slightly dispersed around
the two dominant clusters in Fig.~\ref{fig:sbms}(a), they remain
frequency locked to the collective motion. This is confirmed by
Fig.~\ref{fig:sbms}(b), where both order parameters rapidly converge to
constant values and remain steady in time. Thus, the SBMS is a
stationary synchronized two-cluster state rather than a
standing-wave state. The finite values of both $r_1$ and $r_2$ reflect
the coexistence of first- and second-harmonic coherence arising from the
competition between the two harmonic interactions.

Unlike the standing-wave state observed in bimodal Kuramoto
ensembles \cite{martens2009}, where the two oscillator groups rotate continuously
and the order parameters oscillate periodically in time, the
SBMS is a stationary synchronized state. The order parameters
remain constant in time, indicating that all oscillators are
frequency locked despite the small spread in their phase
distribution. Consequently, the SBMS consists of two stationary
phase-locked clusters rather than counter-rotating oscillator
groups. It may therefore be regarded as the frequency-weighted
unimodal counterpart of the stationary two-cluster synchronized
state reported in biharmonically coupled oscillator populations \cite{komarov2013multiplicity}.

The IC, APMS, and SBMS can be clearly distinguished by the corresponding values of the Daido order parameters. Specifically, the IC is characterized by $r_1\approx0$ and $r_2\approx0$, the APMS by $r_1\approx0$ and $r_2>0$, and the SBMS by finite values of both $r_1$ and $r_2$. These order parameters therefore serve as convenient indicators for identifying the collective state of the system. In the following subsection, we employ one-parameter continuation by varying the coupling strengths to investigate the transitions between these collective states and determine their regions of existence and stability.

\FloatBarrier

\subsection{One-parameter continuation}\label{subsec:continuation}

\begin{figure}[tb]
	\centering
	\includegraphics[width=\linewidth]{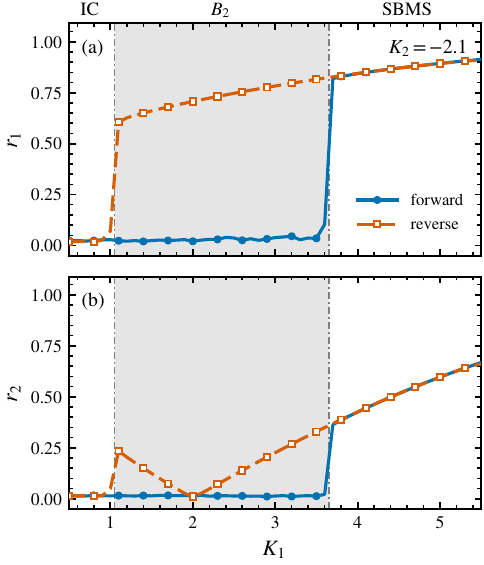}
	\caption{One-parameter continuation in $K_1$ at fixed $K_2=-2.1$, showing (a) the first-harmonic order parameter $r_1$ and (b) the second-harmonic order parameter $r_2$ for the forward (solid) and reverse (dashed) continuations. The shaded region, bounded by the two vertical dashed lines, denotes the bistable region $B_2$, where the incoherent state (IC) and the symmetry-broken multibranch state (SBMS) coexist. The reverse branch of $r_2$ exhibits a re-entrant behavior, reaching a minimum near $K_1\simeq2$ due to the cancellation of the second-harmonic contributions from the two synchronized clusters.}
	\label{fig:sweep_K2_neg2p1}
\end{figure}

To determine the boundaries between the collective states and identify the nature of the corresponding transitions, we perform one-parameter continuation by adiabatically varying one coupling strength while keeping the other fixed. At each continuation step, the asymptotic state obtained for the previous parameter value is used as the initial condition for the next, enabling the system to remain on the same stable solution branch until that branch loses stability. The continuation is performed in both the forward and reverse directions. A difference between the two continuation paths indicates hysteresis and therefore a discontinuous (first-order) transition.

We first examine the transition from the incoherent state (IC) to the symmetry-broken multibranch state (SBMS) by fixing $K_2=-2.1$, for which the second-harmonic interaction is repulsive and the APMS is not supported, and adiabatically varying $K_1$. The corresponding continuation is shown in Fig.~\ref{fig:sweep_K2_neg2p1}. During the forward sweep, the system remains in the incoherent state, with both order parameters close to zero, until the coupling reaches the critical value $K_1\simeq3.65$. At this point, both $r_1$ and $r_2$ increase discontinuously, indicating an abrupt transition to the SBMS. Beyond the transition, both order parameters increase smoothly with increasing $K_1$.

The reverse sweep initially retraces the forward branch but deviates
within the bistable region. Starting from the SBMS, the synchronized state remains stable even below the forward transition point and persists down to $K_1\simeq1$, where it abruptly collapses to the incoherent state. Consequently, the forward and reverse continuations do not coincide over a broad interval of $K_1$, giving rise to the bistable region $B_2$, where the IC and SBMS coexist as stable attractors. The pronounced hysteresis loop clearly demonstrates that the IC--SBMS transition is discontinuous (first-order), consistent with the explosive synchronization induced by frequency-weighted coupling~\cite{xu2016synchronization}.

An interesting feature of the reverse continuation is the non-monotonic variation of the second-order parameter. As $K_1$ decreases within the bistable region, $r_2$ initially decreases and nearly vanishes around $K_1\simeq2$, while $r_1$ remains finite throughout the synchronized branch. Upon further decreasing $K_1$, $r_2$ increases again before the SBMS finally loses stability [Fig.~\ref{fig:sweep_K2_neg2p1}(b)]. This re-entrant behavior does not indicate a loss of synchronization. Instead, it originates from the continuous variation of the phase separation between the two synchronized clusters. As the cluster separation passes through the value corresponding to $\theta=\pi/4$, the second-harmonic contributions from the two clusters cancel each other, causing $r_2=\cos2\theta$ to vanish, whereas the first-order parameter $r_1=\cos\theta$ remains finite. This behavior follows directly from the two-cluster relations given in Eq.~(\ref{eq:ops_two_cluster}) of Sec.~\ref{sec:sbms_conditions}.

\begin{figure}[tb]
\centering
\includegraphics[width=\linewidth]{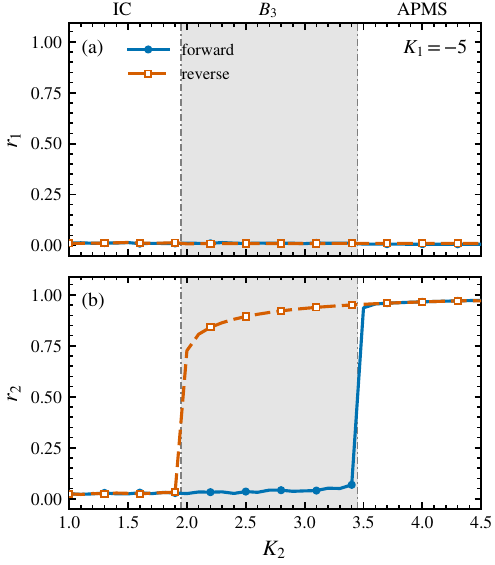}
\caption{One-parameter continuation in $K_2$ at fixed $K_1=-5$, showing
(a) the first-harmonic order parameter $r_1$ and (b) the second-harmonic
order parameter $r_2$ for the forward (solid) and reverse (dashed)
continuations. While $r_1$ remains close to zero throughout the
continuation, $r_2$ exhibits a pronounced hysteresis loop associated
with the transition between the incoherent state (IC) and the antipodal
multibranch state (APMS). The shaded region, bounded by the two vertical
dashed lines, denotes the bistable region $B_3$, where the IC and APMS
coexist.}
\label{fig:sweep_K1_neg5}
\end{figure}

Next, we investigate the transition from the incoherent state (IC) to the
antipodal multibranch state (APMS) by fixing $K_1=-5$, for which the
first-harmonic interaction is repulsive, and adiabatically varying
$K_2$. The corresponding continuation is shown in
Fig.~\ref{fig:sweep_K1_neg5}. Throughout the continuation, the first-harmonic order parameter remains close to zero for both the forward and reverse sweeps [Fig.~\ref{fig:sweep_K1_neg5}(a)]. In contrast, the second-harmonic order parameter becomes finite along the synchronized branch [Fig.~\ref{fig:sweep_K1_neg5}(b)], confirming that the ordered state is the APMS. Consequently, the antipodal arrangement of the synchronized clusters causes the first-harmonic contributions from the two clusters to cancel, resulting in a vanishing first-harmonic order parameter.

The second-harmonic order parameter exhibits a pronounced hysteresis loop
[Fig.~\ref{fig:sweep_K1_neg5}(b)]. During the forward continuation, the
system remains in the incoherent state until the coupling reaches the
critical value $K_2\simeq3.5$, where $r_2$ increases discontinuously,
signalling the abrupt emergence of the APMS. Beyond this point, $r_2$
increases smoothly with increasing $K_2$. During the reverse
continuation, however, the APMS remains stable well below the forward
transition point and persists down to $K_2=2$, where it abruptly
collapses to the incoherent state. Consequently, the forward and reverse
continuations enclose the bistable region $B_3$, in which the IC and
APMS coexist as stable attractors. The resulting hysteresis loop confirms
that the IC--APMS transition is discontinuous (first-order), consistent
with the explosive synchronization induced by the frequency-weighted
coupling~\cite{xu2016synchronization}.

The termination of the reverse branch of the second-harmonic order
parameter in Fig.~\ref{fig:sweep_K1_neg5}(b) provides a direct
verification of the analytical self-consistency relation,
Eq.~\eqref{eq:r2_apms}. The theory predicts that the antipodal branch
ceases to exist at $K_2=2$ with $r_2=1/\sqrt{2}$. Both predictions are
in good agreement with the numerical continuation: the APMS loses
stability precisely at $K_2=2$, and the corresponding value of the
second-harmonic order parameter agrees with $r_2=1/\sqrt{2}$. This
agreement provides strong support for the analytical description of the
APMS.

\begin{figure}[tb]
	\centering
	\includegraphics[width=\linewidth]{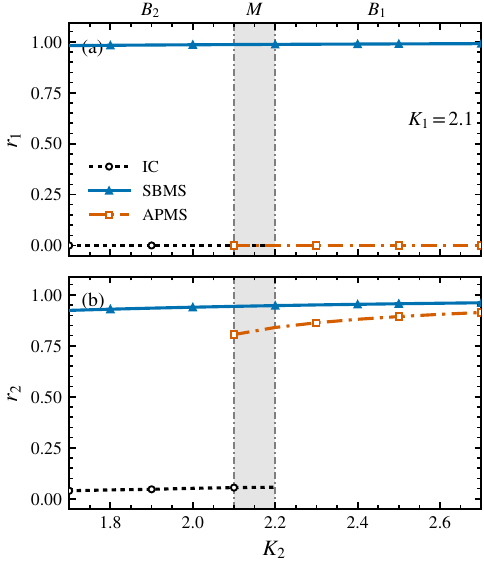}
	\caption{Branch continuation in $K_2$ at fixed $K_1=2.1$, showing
(a) the first-harmonic order parameter $r_1$ and (b) the second-harmonic
order parameter $r_2$. The IC, APMS, and SBMS branches are obtained by
independently continuing each stable state. The IC and APMS branches
terminate at $K_2=2.2$ and $K_2=2.1$, respectively, whereas the SBMS
remains stable throughout. The shaded region $M$ denotes the tristable
region, where the IC, APMS, and SBMS coexist as stable attractors.}
	\label{fig:sweep_K1_2p1}
\end{figure}

Finally, we demonstrate the existence of a tristable region by fixing
$K_1=2.1$ and adiabatically varying $K_2$. This value of $K_1$ lies within
the bistable region $B_2$, where the incoherent state (IC) and the
symmetry-broken multibranch state (SBMS) coexist. Accordingly, two
different initial conditions are employed: one converging to the IC and
the other to the SBMS. The corresponding continuations are shown in
Fig.~\ref{fig:sweep_K1_2p1}.

Starting from the SBMS initial condition, the synchronized branch remains
stable throughout the continuations, indicating
that the SBMS persists over the entire range of $K_2$ considered. In
contrast, the branch initiated from the IC remains incoherent only up to
a critical value of $K_2\simeq2.2$, where it undergoes an abrupt
transition to the antipodal multibranch state (APMS). During the reverse
continuation, the APMS remains stable down to $K_2=2.1$, where it
discontinuously returns to the IC. The difference between the forward and
reverse transition points establishes a hysteresis loop, confirming the
bistability between the IC and APMS.

Since the SBMS remains stable over the entire continuation, the
IC--APMS bistable region overlaps with the pre-existing IC--SBMS
bistable region. Consequently, within the interval
$2\lesssim K_2\lesssim2.2$, all three collective states including the IC, APMS,
and SBMS are simultaneously stable. This overlap defines the tristable
region $M$, indicated by the shaded band in
Fig.~\ref{fig:sweep_K1_2p1}. For $K_2<2$, only the IC and SBMS coexist,
corresponding to the bistable region $B_2$, whereas for
$K_2\gtrsim2.2$, the IC loses stability, leaving the APMS and SBMS as
the two coexisting stable states, which defines the bistable region
$B_1$.

Taken together, Figs.~\ref{fig:sweep_K2_neg2p1},
\ref{fig:sweep_K1_neg5}, and
\ref{fig:sweep_K1_2p1} illustrate the complementary roles of the two
harmonic interactions. Increasing the attractive first-harmonic coupling
at fixed repulsive second-harmonic coupling drives a discontinuous
transition from the IC to the SBMS, where both $r_1$ and $r_2$ are
finite, giving rise to the bistable region $B_2$. Conversely, increasing
the attractive second-harmonic coupling at fixed repulsive
first-harmonic coupling drives a discontinuous transition from the IC to
the APMS, characterized by a finite $r_2$ and $r_1\simeq0$, producing
the bistable region $B_3$. When these two bistable regions overlap, the
IC, APMS, and SBMS coexist, forming the tristable region $M$. Beyond
this overlap, the IC loses stability, leaving the APMS and SBMS as the
two coexisting stable states in the bistable region $B_1$.

\section{Analytical Results}
\label{sec:4}
To complement the numerical results, we develop an analytical
description of the collective dynamics. We first determine the linear
stability threshold of the incoherent state, followed by an analysis of the existence and stability of the symmetry-broken multibranch state (SBMS). Together, these results provide an analytical framework for understanding the phase boundaries and synchronization transitions of the model.
\subsection{Linear stability of the incoherent state}%
\label{sec:ic_stability}

In this section, we determine the critical coupling strengths at which the Incoherent state first becomes linearly unstable. To begin, note that in the continuum description, the incoherent state is represented by the uniform density $f_0(\theta,\omega) = g(\omega)/(2\pi)$, where both order parameters vanish, i.e., $r_1 = r_2 = 0$. This state is a stationary solution of the continuity equation~\eqref{eq:cont_eq} for all couplings. This is because, with $Z_1 = Z_2 = 0$, the velocity reduces to $v = \omega$, so $\partial_\theta(\omega f_0) = 0$.

To test its stability, we perturb the uniform density,
\begin{equation}\label{eq:pert_ansatz}
  f(\theta,\omega,t) = f_0(\theta,\omega) + \varepsilon\,
    h(\theta,\omega,t),
  \qquad 0 < \varepsilon \ll 1,
\end{equation}
so that the order parameters acquire corrections $O(\varepsilon)$, 
$Z_m = \varepsilon\,\delta Z_m + O(\varepsilon^2)$, with
$\delta Z_m(t) = \int_{-\infty}^{\infty}\!\int_0^{2\pi} e^{im\theta}\,
h\,d\theta\,d\omega$.
Substituting Eq.~\eqref{eq:pert_ansatz} into the continuity
equation~\eqref{eq:cont_eq} and collecting terms at $O(\varepsilon)$, and
using the fact that $f_0$ is independent of $\theta$, gives the
linearized equation
\begin{equation}\label{eq:lin_cont}
  \partial_t h + \omega\,\partial_\theta h
    = f_0\,|\omega|\sum_{m=1}^{2}
        \frac{m K_m}{2}
        \bigl(\delta Z_m\, e^{-im\theta}
              + \overline{\delta Z}_m\, e^{im\theta}\bigr).
\end{equation}
The forcing on the right-hand side involves only the harmonics $e^{\pm i\theta}$ and $e^{\pm 2i\theta}$. These harmonics excite only the $m=1$ and $m=2$ sectors. In linear order, perturbations in the $m=1$ sector (driven by $K_1$) are decoupled from those in the $m=2$ sector (driven by $K_2$). This decoupling is due to the structure of the forcing. As a result, the two critical couplings $K_1^{(c)}$ and $K_2^{(c)}$ can be computed independently.

We therefore fix a single harmonic sector $m \in \{1,2\}$ and seek a
normal mode
\begin{equation}\label{eq:normal_mode}
  h(\theta,\omega,t) = b(\omega)\, e^{\lambda t}\, e^{-im\theta}
    + \text{c.c.},
\end{equation}
where $\lambda \in \mathbb{C}$ is the growth rate and ``c.c.'' denotes the
complex conjugate, ensuring that $h$ is real. For this mode
$\partial_t h = \lambda h$ and $\partial_\theta h = -im\,h$, so that
projecting Eq.~\eqref{eq:lin_cont} onto $e^{-im\theta}$ yields
\begin{equation}\label{eq:bw}
  (\lambda - im\omega)\, b(\omega)
    = \frac{g(\omega)}{2\pi}\,|\omega|\,\frac{m K_m}{2}\,\delta Z_m.
\end{equation}
Substituting the normal mode into the definition
of $\delta Z_m$ gives the self-consistency relation
$\delta Z_m = 2\pi \int_{-\infty}^{\infty} b(\omega)\,d\omega$. Solving
Eq.~\eqref{eq:bw} for $b(\omega)$, inserting it into this relation, and
cancelling the nonzero factor $\delta Z_m$ produces the dispersion
relation
\begin{equation}\label{eq:dispersion}
  1 = \frac{m K_m}{2}\int_{-\infty}^{\infty}
        \frac{|\omega|\,g(\omega)}{\lambda - im\omega}\,d\omega.
\end{equation}
The incoherent state loses stability in sector $m$ when a root $\lambda$
of Eq.~\eqref{eq:dispersion} crosses into $\operatorname{Re}(\lambda)>0$.

Because of the frequency-weighting, the crossing typically occurs
at a nonzero frequency. We therefore set $\lambda \to 0^+ + i\Omega$ with
$\Omega \in \mathbb{R}$ and define
\begin{equation}\label{eq:Im_def}
  \mathcal{I}_m(\Omega)
    = \int_{-\infty}^{\infty}
         \frac{|\omega|\,g(\omega)}{0^+ + i(\Omega - m\omega)}\,d\omega,
\end{equation}
so that Eq.~\eqref{eq:dispersion} reads $1 = (m K_m/2)\,\mathcal{I}_m(\Omega)$.
Using the standard distributional identity
$1/(x \pm i0^+) = \text{P.V.}(1/x) \mp i\pi\delta(x)$, the integrand of
Eq.~\eqref{eq:Im_def} splits into a real (delta) part and an imaginary
(principal-value) part,
\begin{equation}\label{eq:Im_split}
\begin{split}
  \mathcal{I}_m(\Omega)
    = {}& \pi \int_{-\infty}^{\infty}
        |\omega|\,g(\omega)\,\delta(\Omega-m\omega)\,d\omega \\
    &- i\,\text{P.V.}\!\int_{-\infty}^{\infty}
        \frac{|\omega|\,g(\omega)}{\Omega - m\omega}\,d\omega.
\end{split}
\end{equation}
Since the left-hand side of the dispersion relation is real, the
imaginary part of $\mathcal{I}_m$ must vanish, which fixes the marginal
frequency $\Omega$ through
\begin{equation}\label{eq:pv_cond}
  \text{P.V.}\!\int_{-\infty}^{\infty}
    \frac{|\omega|\,g(\omega)}{\Omega - m\omega}\,d\omega = 0,
\end{equation}
while the real part fixes the critical coupling through
$1 = (m K_m/2)\,\operatorname{Re}\mathcal{I}_m(\Omega)$.

We now evaluate the principal-value condition~\eqref{eq:pv_cond} for the
Lorentzian~\eqref{eq:lorentz}. The weight
$W(\omega) \equiv |\omega|\,g(\omega)$ is an even function of $\omega$.
Splitting the integral and substituting $\omega \to -\omega$ in the
negative half-line combines the two contributions, so that for
$\Omega \neq 0$,
\begin{equation}\label{eq:pv_reduced}
  \text{P.V.}\!\int_{-\infty}^{\infty}
    \frac{W(\omega)}{\Omega - m\omega}\,d\omega
  = 2\Omega\,\text{P.V.}\!\int_0^{\infty}
      \frac{\omega\,g(\omega)}{\Omega^2 - m^2\omega^2}\,d\omega.
\end{equation}
Substituting
$u = \omega^2$ and applying the partial-fraction decomposition
\begin{equation}\label{eq:pf}
  \frac{1}{(1+u)(\Omega^2 - m^2 u)}
    = \frac{1}{\Omega^2+m^2}
      \left(\frac{1}{1+u}
            + \frac{m^2}{\Omega^2 - m^2 u}\right),
\end{equation}
the principal-value integral evaluates in closed form to
\begin{equation}\label{eq:pv_result}
  \text{P.V.}\!\int_{-\infty}^{\infty}
    \frac{|\omega|\,g(\omega)}{\Omega - m\omega}\,d\omega
  = \frac{\Omega}{\pi(\Omega^2+m^2)}\,
    \ln\!\frac{\Omega^2}{m^2}.
\end{equation}
Setting Eq.~\eqref{eq:pv_result} to zero for $\Omega \neq 0$ forces
$\ln(\Omega^2/m^2)=0$, that is,
\begin{equation}\label{eq:hopf_freq}
  \Omega = \pm m.
\end{equation}
The incoherent state therefore loses stability through a Hopf-type
crossing at frequency $\Omega = m$ in the $m$th harmonic sector.

It remains to evaluate the real part of $\mathcal{I}_m$. The
delta function $\delta(\Omega - m\omega)$ is supported at
$\omega_0 = \Omega/m$, and the scaling rule
$\delta(\Omega - m\omega) = |m|^{-1}\delta(\omega-\omega_0)$ gives
$\operatorname{Re}\mathcal{I}_m(\Omega) = \pi\,|\omega_0|\,g(\omega_0)/m$.
At the Hopf frequency~\eqref{eq:hopf_freq} we have $\omega_0 = 1$, so that
$|\omega_0| = 1$ and $g(1) = 1/[\pi(1+1)] = 1/(2\pi)$, where
$\operatorname{Re}\mathcal{I}_m = 1/(2m)$. Inserting this into the
real-part condition gives
\begin{equation}\label{eq:Kmc}
  1 = \frac{m K_m}{2}\cdot\frac{1}{2m} = \frac{K_m}{4}
  \qquad\Longrightarrow\qquad
  K_m^{(c)} = 4.
\end{equation}
Because the two harmonic sectors decouple at linear order, the stability
thresholds for the first and second harmonics are obtained
independently,
\begin{equation}\label{eq:as_thresholds}
  K_1^{(c)} = 4,\quad \Omega_1 = 1;
  \qquad
  K_2^{(c)} = 4,\quad \Omega_2 = 2.
\end{equation}
In the $(K_2,K_1)$ parameter plane, the incoherent state is therefore
linearly stable for $K_1 < 4$ and $K_2 < 4$ simultaneously. The boundary
consists of two straight half-lines, $K_1 = 4$ for $K_2 < 4$ and
$K_2 = 4$ for $K_1 < 4$, meeting at the corner point
$(K_2, K_1) = (4, 4)$.

\subsection{Existence and local stability of the SBMS}%
\label{sec:sbms_conditions}

The SBMS is approximated by a locked two-cluster ansatz in which oscillators
with $\sigma=+1$ lock at phase $+\theta$ and oscillators with $\sigma=-1$
lock at phase $-\theta$, with $0<\theta<\pi/2$. The phase separation between
the two clusters is therefore $\Delta\theta=2\theta$, and symmetry breaking
corresponds to $\Delta\theta\neq\pi$.

For this two-cluster state, the Daido order parameters~\eqref{eq:daido}
reduce to
\begin{equation}\label{eq:ops_two_cluster}
  r_1=\cos\theta, \qquad r_2=\cos2\theta .
\end{equation}
Substituting Eq.~\eqref{eq:ops_two_cluster} into the locking
condition~\eqref{eq:lock} for the $\sigma=+1$ cluster gives
\begin{equation}\label{eq:sbms_exist}
  1=K_1\cos\theta\sin\theta
     +K_2\cos2\theta\sin2\theta .
\end{equation}
Using $2\sin\theta\cos\theta=\sin2\theta$ and
$2\sin2\theta\cos2\theta=\sin4\theta$, this becomes
\begin{equation}\label{eq:sbms_exist_compact}
  1=\frac{K_1}{2}\sin2\theta
    +\frac{K_2}{2}\sin4\theta .
\end{equation}
Solving for the first-harmonic coupling gives the SBMS existence relation
\begin{equation}\label{eq:K1_exist}
  K_1(\theta;K_2)=\frac{2}{\sin2\theta}-2K_2\cos2\theta .
\end{equation}
Thus, for a prescribed $K_2$, each admissible value of $\theta$ specifies the
value of $K_1$ required to support a locked symmetry-broken two-cluster
state with separation $2\theta$.

Existence alone is insufficient: the locked phases must also be stable to
small perturbations. Writing the gauge-fixed velocity~\eqref{eq:vfield_gauge}
as $\dot\theta=|\omega|F(\theta)$ with
\begin{equation}
  F(\theta)=\sigma-K_1r_1\sin\theta-K_2r_2\sin2\theta,
\end{equation}

a fixed point is locally stable when $F'(\theta)<0$. Differentiation and use
of Eq.~\eqref{eq:ops_two_cluster} give
\begin{equation}\label{eq:Fprime}
  F'(\theta)=-\left(K_1\cos^2\theta+2K_2\cos^2 2\theta\right),
\end{equation}
so the local-stability condition is
\begin{equation}\label{eq:stab_cond}
  K_1\cos^2\theta+2K_2\cos^2 2\theta>0 .
\end{equation}

The two terms in Eq.~\eqref{eq:stab_cond} are the restoring stiffnesses due
to the first and second harmonics. A positive second harmonic contributes to
stability, whereas a negative second harmonic can reduce the net stiffness
and exclude otherwise existing locked solutions.

The lower SBMS boundary used in the phase diagram is obtained by minimizing
the existence relation over the stable locked solutions,
\begin{equation}\label{eq:K1c_stable_min}
\begin{split}
  K_{1,c}(K_2)={}&\min_{0<\theta<\pi/2}
  \bigl\{K_1(\theta;K_2):\\
  &\qquad K_1\cos^2\theta+2K_2\cos^2 2\theta>0\bigr\} .
\end{split}
\end{equation}
When the minimum occurs in the interior of the stable interval, it is fixed
by $\partial K_1/\partial\theta=0$, which gives
\begin{equation}\label{eq:fold_implicit}
  \cos2\theta=K_2\sin^3 2\theta .
\end{equation}
In particular, for $K_2=0$ the minimizing angle is $\theta=\pi/4$, yielding
$K_{1,c}(0)=2$. For $K_2=2$, the minimizing angle is $\theta=\pi/8$, giving
\begin{equation}\label{eq:K1c_at_2}
  K_{1,c}(2)
  =\frac{2}{\sin(\pi/4)}-2(2)\cos(\pi/4)=0 .
\end{equation}
Equations~\eqref{eq:K1_exist}, \eqref{eq:stab_cond}, and
\eqref{eq:K1c_stable_min} provide the analytical criterion used to compare
with the lower SBMS boundary extracted from the numerical phase diagram.

\FloatBarrier


\section{Coupling Parameter space}\label{subsec:pd_overview}

\begin{figure}[tb]
	\centering
	\includegraphics[width=\linewidth]{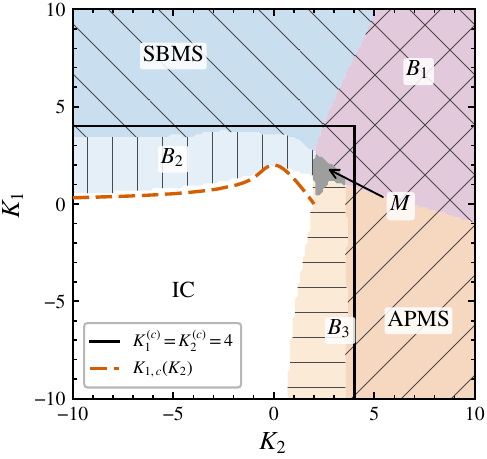}
	\caption{Phase diagram of model~\eqref{eq:model_full} in the $(K_2,K_1)$
		plane, obtained by classifying the state reached from three prepared
		initial conditions at every grid point. Seven regions are
		distinguished by fill and hatch: the three states IC, APMS and SBMS,
		the bistable regions $B_1$ (APMS--SBMS), $B_2$ (IC--SBMS) and $B_3$
		(IC--APMS), and the multistable region $M$ (grey), in which all three
		states are stable together. The solid lines are the linear
		instability thresholds $K_1^{(c)}=K_2^{(c)}=4$ of the incoherent
		state and the dashed curve is the lower SBMS boundary
		$K_{1,c}(K_2)$ of Eq.~\eqref{eq:K1c_stable_min}.}
	\label{fig:phase_diagram}
\end{figure}

We now assemble the one-parameter continuation results of
Sec.~\ref{subsec:continuation} into a global phase diagram by exploring
the $(K_2,K_1)$ parameter plane. At each grid point, three
independently prepared initial conditions corresponding to the IC, APMS,
and SBMS are evolved, and the resulting asymptotic states are classified.
The resulting phase diagram is shown in
Fig.~\ref{fig:phase_diagram}. Seven regions are identified: the
single-state regions IC, APMS, and SBMS; the bistable regions
$B_1$ (APMS--SBMS), $B_2$ (IC--SBMS), and $B_3$ (IC--APMS); and the
tristable region $M$, where all three collective states coexist as
stable attractors.

The overall organization of the phase diagram reflects the competition
between the first- and second-harmonic interactions. The incoherent
state occupies the lower-left region, where neither coupling is
sufficiently strong to establish synchronization. The APMS dominates
when the attractive second harmonic is sufficiently strong, whereas the
SBMS occupies the region where the attractive first harmonic prevails.
Separating these single-state regions are the bistable regions $B_1$,
$B_2$, and $B_3$, whose finite widths are a consequence of the
subcritical nature of the corresponding synchronization transitions.

The boundaries of the incoherent region are governed by the
linear-instability thresholds $K_1^{(c)}=K_2^{(c)}=4$, derived in
Sec.~\ref{sec:ic_stability} and indicated by the solid lines in
Fig.~\ref{fig:phase_diagram}. Near the coordinate axes these analytical
thresholds accurately delimit the IC region and cap the bistable windows
$B_2$ and $B_3$. However, this agreement deteriorates near the diagonal
$K_1\approx K_2$, where both harmonic interactions act simultaneously.
Since both synchronization transitions are subcritical, the ordered
states already exist as finite-amplitude attractors before the linear
stability thresholds are reached. Consequently, the incoherent state is
replaced by the synchronized states within a region where linear theory
still predicts stability, causing the numerically observed IC boundary
to lie well inside the corner bounded by $K_1=4$ and $K_2=4$.

The dashed curve in Fig.~\ref{fig:phase_diagram} represents the
analytical lower stability boundary of the SBMS,
$K_{1,c}(K_2)$, given by
Eq.~\eqref{eq:K1c_stable_min}. As discussed in
Sec.~\ref{sec:sbms_conditions}, this boundary is governed by two
different mechanisms. For moderate negative values of $K_2$, it is
determined by the saddle-node (fold) condition of the existence
relation, whereas for larger negative $K_2$ it is controlled by the
local stability criterion, with the repulsive second harmonic reducing
the restoring force of the locked state. The branch continuation shown
in Fig.~\ref{fig:sweep_K2_neg2p1} intersects this boundary, and the
termination of the reverse SBMS branch coincides with the analytical
prediction.

The upper-right quadrant contains the bistable region $B_1$, where the
APMS and SBMS coexist as stable attractors. This region is consistent
with the branch continuation results of
Fig.~\ref{fig:sweep_K1_2p1}, where the two ordered states remain stable
over a broad interval of $K_2$. Compared with the relatively narrow
bistable regions $B_2$ and $B_3$, the wider extent of $B_1$ indicates
that the competition between two synchronized states is considerably
more robust than the competition between an ordered and the incoherent
state.

Finally, the tristable region $M$ appears near the intersection of the
three bistable regions $B_1$, $B_2$, and $B_3$. In this region the IC,
APMS, and SBMS are simultaneously stable, and the observed asymptotic
state depends entirely on the initial condition. The existence of this
region is confirmed by the branch continuation of
Fig.~\ref{fig:sweep_K1_2p1}, which demonstrates the overlap of the
IC--SBMS and IC--APMS bistable regions while the SBMS remains stable
throughout. Thus, the phase diagram provides a unified picture of the
competition between the two harmonic interactions and the resulting
hierarchy of mono-, bi-, and tristable dynamical regimes.

\FloatBarrier

\section{Conclusions}\label{sec:conclusion}

We have investigated the collective dynamics of globally coupled phase
oscillators with frequency-weighted biharmonic interactions, which
combine frequency-weighted coupling and biharmonic interactions within a
unified Kuramoto framework. The interplay between frequency weighting
and the competing first- and second-harmonic interactions gives rise to
a rich phase diagram in the $(K_2,K_1)$ parameter plane. The diagram
contains three collective states: the incoherent state (IC), the
antipodal multibranch state (APMS), and the symmetry-broken multibranch
state (SBMS). It further exhibits three bistable regions and a narrow
tristable region where all three states coexist as stable attractors.

The APMS is characterized by antipodal multibranch phase locking with
$r_1\simeq0$ and finite $r_2$, whereas the SBMS consists of two
phase-locked clusters separated by a phase difference
$\Delta\theta\neq\pi$, resulting in finite values of both $r_1$ and
$r_2$. Although the governing equations retain the global rotational
symmetry of the Kuramoto model and, owing to the symmetric frequency
distribution, are also invariant under the reflection transformation
$(\theta,\omega)\rightarrow(-\theta,-\omega)$, the synchronized
solutions organize these symmetries in fundamentally different ways.
The APMS preserves the reflection symmetry through its antipodal
($\pi$-periodic) organization promoted by the second-harmonic
interaction, whereas the SBMS emerges through a spontaneous breaking of
this reflection symmetry, producing an asymmetric two-cluster
configuration. Thus, the transition from the APMS to the SBMS may be
viewed as a spontaneous symmetry-breaking transition between two ordered
collective states.

Analytically, we derived the linear instability thresholds of the
incoherent state together with the self-consistency condition for the
APMS and the existence and local stability conditions for the SBMS.
These analytical predictions accurately determine the principal phase
boundaries and are in excellent agreement with extensive numerical
simulations. The phase diagram further reveals how the competition
between the first- and second-harmonic interactions generates three
distinct bistable regions and, through the overlap of the IC--SBMS and
IC--APMS bistabilities, a narrow tristable region in which the IC,
APMS, and SBMS coexist.

The present work demonstrates that the combination of frequency
weighting and biharmonic coupling generates collective phenomena that
are absent when either mechanism acts in isolation. In particular, the
competition between a reflection-symmetric antipodal synchronized state
and a spontaneously symmetry-broken synchronized state organizes the
collective dynamics into mono-, bi-, and tristable regimes. These
results establish a direct connection between symmetry, spontaneous
symmetry breaking, multistability, and explosive synchronization in
frequency-weighted oscillator networks, while providing a new mechanism
for realizing antipodal synchronization in unimodal oscillator
populations without requiring multimodal frequency distributions.

\begin{acknowledgments}
The work of R.G. and V.K.C. was supported by the DST-SERB Core Research Grant (CRG) under Grant No.~CRG/2023/003505 and the ANRF Advanced Research Grant (ARG) under Grant No.~ANRF/ARG/2025/004108/PS. R.S. thanks Dr.~Gourab Kumar Sar (University of Calgary, Canada) for valuable discussions and helpful literature suggestions.
\end{acknowledgments}

\section*{Author Declarations}

\subsection*{Conflict of Interest}
The authors have no conflicts to disclose.

\subsection*{Author Contributions}
\textbf{R.~Senthamizhan}: Conceptualization (equal); Investigation
(lead); Software (lead); Writing -- original draft (lead).
\textbf{R.~Gopal}: Conceptualization (equal); Supervision (equal);
Writing -- review \& editing (equal).
\textbf{V.~K.~Chandrasekar}: Conceptualization (equal);
 Supervision (equal);
Writing -- review \& editing (equal).

\section*{Data Availability Statement}
The data that support the findings of this study are available
from the corresponding author upon reasonable request.

\bibliography{references}

\end{document}